\documentclass[doublecol]{epl2} 

\usepackage{graphicx}
\usepackage{amsmath} 
\usepackage[colorlinks]{hyperref}

\newcommand{\mb}[1]{\mathbf{#1}} \newcommand{\mr}[1]{\mathrm{#1}}

\title{The end points in the dispersion of Holstein polarons}
\author{Glen L. Goodvin \and Mona Berciu}
\shortauthor{G. L. Goodvin \etal}

\institute{Department of Physics \& Astronomy, University of British
 Columbia, Vancouver, BC, Canada V6T 1Z1 
}


\abstract{We investigate the existence of end points in the dispersion
 of Holstein polarons in various dimensions, using the Momentum
 Average approximation which has proved to be very accurate for
 this model. An end point separates momenta for which the
 lowest-energy state is a discrete level, {\em i.e.}, an
 infinitely-lived polaron, from those where the lowest-energy feature
 is a continuum in which the ``polaron'' is signalled by a resonance
 with a finite
 lifetime. While such end
 points are known to not appear in 1D, we show here that they are
 generic in 3D if the particle-boson coupling is not too
 strong. The 2D case is ``critical'': a pure 2D Holstein model
 has no end points, like the 1D case. However, any amount of
 interlayer hopping leads to 3D-like behavior. As a result, such end
 points are expected to appear in the spectra of layered, quasi-2D
 systems described by Holstein models. Generalizations to other
 models are also briefly discussed.
}

\begin{document}

\maketitle

%%%%%%%%%%%%%%%%%%%%%%%%%%%%%%%%%%%
\section{Introduction}

%BACKGROUND: 
Understanding the properties of a novel material is inherently linked
to understanding the
physics of its underlying quasi-particles. The formation of such
quasi-particles is a very old problem that comes up again and again in
condensed matter physics. In particular, the formation of a
quasi-particle \emph{qp} due to the coupling of the particle to
bosonic degrees of freedom is relevant in many systems of current
interest, such as manganites, cuprates, most other oxides, various organic
materials, etc \cite{salamon:01,shen:04,hengsberger:99,matsui:10}. This coupling can be to phonons, magnons, orbitons, or any combination thereof, and the resulting
composite object is known as a polaron.
%

%OBJECTIVES: 
Such polaron problems continue to attract considerable analytical and
numerical interest, particularly for simple lattice Hamiltonians like the
Holstein model \cite{holstein:59} which can be studied by many 
means \cite{fehske:07}. The
ground-state is a discrete eigenstate, \emph{i.e.}, it describes
a true quasiparticle with an infinite lifetime. At higher energies, however, the
eigenstates may describe incoherent scattering of the particle on bosonic
excitations, and have finite lifetimes. 

One important question which has not been answered for lattice models of
polarons, and which we settle for the Holstein polaron in this Letter, is whether the
lowest eigenstate for a given momentum is an infinitely-lived,
discrete state throughout the Brillouin zone, or not. That such a
state may not exist for all momenta has long been suspected, since 
the dispersion of the Fr\"ohlich polaron exhibits a so-called {\em end
 point} \cite{Prok}. While this is a three-dimensional, continuous
model, it is interesting to see if similar phenomenology is possible
for lattice models, and under what conditions.

As we show below, the answer depends strongly on both the
dimensionality of the system, and the value of the effective
particle-boson coupling. At first sight, the former seems surprising,
since many authors have demonstrated that dimensionality plays a very
small role in determining the qualitative features of the ground state
properties of the Holstein polaron~\cite{romero:99,ku:02, berciu:06}. However,
the bare electron and polaron densities of states vary significantly with
dimension \cite{kornilovich:99}, and as we show below, this plays an essential role in
deciding whether a discrete level is pushed below the
polaron+one-phonon continuum, or not. This continuum appears
at $\Omega$ (the energy of the boson, modelled as an Einstein mode)
above the ground-state energy, and describes excited states with a
boson far from the polaron.

That the effective coupling is also very important is, on the other
hand, not
surprising. At very large couplings, the polaron effective mass is
exponentially large, and therefore the polaron dispersion is very
flat. As a result, one expects that the entire polaron band fits below
the continuum, {\em i.e.}, a true quasiparticle state exists at low
energies in the entire Brillouin zone. At weaker couplings, however,
the polaron dispersion is considerable and its band may overlap with
the 
continuum at larger momenta. For the most studied case --
the 1D chain -- it is well known that irrespective of the coupling,
(at least) one discrete state is pushed below the continuum everywhere
in the Brillouin zone, although it flattens out just underneath the
continuum at larger momenta. While it seems to be assumed in the literature
that this is also the case in higher dimensions, we show here that
this is not true in 3D. Indeed, in 3D at weak couplings, an infinitely
lived polaron exists only in a finite region near the $\Gamma$ point.
The 2D case is marginal: while a low-energy discrete
state always exists in a purely 2D model, any amount of anisotropic
hopping in the 3rd direction renders the system effectively 3D-like,
and the transition from a true quasiparticle to a resonance becomes
possible if the coupling is not too strong. In other words, unlike 1D
systems, layered
quasi-2D systems are not guaranteed to have a long-lived 
quasiparticle as the low-energy eigenstate throughout the Brillouin
zone.

\section{Formalism}

%Review/Summary of MA
Our results are based on the Momentum Average (MA) approximation,
initially developed to study the properties of the Holstein
model~\cite{berciu:06} and then extended to more complex polaron
models~\cite{goodvin:08}. The 
approximation allows one to sum \emph{all} diagrams appearing in the
expansion of the self-energy, albeit with exponentially small
contributions discarded so 
that this sum can actually be performed analytically. The MA
approximation has been shown to be accurate over all of
parameter space (except the extremely adiabatic limit), in all
dimensions, and for all energies. Furthermore, 
the approximation satisfies the first six spectral weight sum rules
exactly, and all higher order sum rules to a good degree of
accuracy. It can also be systematically improved by including
additional states, allowing for the correct reproduction of the
polaron+one-phonon continuum. These improvements also lead to increased
accuracy throughout parameter space, and exact agreement with additional
spectral weight sum rules. Recently, the MA approximation has also been
generalized to treat broken translational symmetry, due to
disorder or surfaces~\cite{berciu:10}, and also to obtain
two-particle Green's functions needed for 
a calculation of the optical conductivity~\cite{goodvin:10}. In
conclusion, the MA approximation is a powerful and well-understood
tool 
that allows us to understand very accurately the Holstein polaron
physics, throughout the parameter space, for all energies and
momenta. Also, because it is an analytical approximation, we can
easily probe large regions of parameters space that are difficult
for numerical techniques to investigate, especially in higher dimensions.

%Model
The Holstein model that we analyze here describes the simplest
possible electron-phonon coupling on a lattice, and is given by ~\cite{holstein:59}: 
\begin{equation}
\nonumber
{\cal H} = \sum_{\mb{k}} \left(
\varepsilon^{}_{\mb{k}}c^{\dagger}_{\mb{k}}c^{}_{\mb{k}} + \Omega
b^{\dagger}_{\mb{k}} b^{}_{\mb{k}} \right) + \frac{g}{\sqrt{N}}
\sum_{\mb{k}, \mb{q}} c^{\dagger}_{\mb{k} - \mb{q}}c^{}_{\mb{k}}
\left( b^{\dagger}_{\mb{q}} + b^{}_{-\mb{q}} \right). 
\end{equation}
The first term is the kinetic energy of the free electron, with
$c_{\mb{k}}^{\dagger}$ being electron creation operators for a state
with momentum $\mb{k}$ (the spin is 
irrelevant in this model, and we ignore it in the following). For the
free electron dispersion, we use 
nearest-neighbor hopping on a $d$-dimensional simple hypercubic lattice of
lattice constant $a$: 
\begin{equation}
\varepsilon_{\mb{k}} = -2t\sum_{i=1}^d \cos(k_i a).
\end{equation}
The coupling is to a branch of Einstein optical phonons of frequency
$\Omega$, where $b_{\mb{k}}^{\dagger}$ and $b^{}_{\mb{k}}$ are the
phonon creation and annihilation operators. The last term describes a
momentum-independent on-site linear coupling $g\sum_i c_i^{\dagger}
c_i^{}(b_i^{\dagger} + b_i^{})$, written in $\mb{k}$-space. All sums
over momenta are over the first Brillouin zone. For all results shown
here, the total number of sites $N$ is taken to be infinity. We also
set $\hbar=1$ and 
$a=1$ throughout. 

%MA Result
The MA approximation has been applied to the Holstein model
previously, and its meaning and accuracy is very well
understood \cite{berciu:06}. Although the MA$^{(0)}$ level of approximation has been
shown to be quite accurate at describing ground state properties, it
fails to reproduce the correct location for the
polaron+one-phonon continuum. At the MA$^{(1)}$ level and beyond,
this continuum is accurately reproduced. Although the discussion and
conclusions below do not qualitatively depend on the precise location
of this continuum, we will employ the more accurate MA$^{(1)}$
approximation throughout this work so that our results are
quantitatively, not just qualitatively, accurate.

Application of the MA approximation to the Holstein Hamiltonian allows
us to calculate the single-particle (retarded) Green's function,
defined as $G(\mb{k}, \omega) = \langle 0 | c^{}_{\mb{k}} (\omega -
{\cal H} + i\eta)^{-1} c^{\dag}_{\mb{k}}| 0
\rangle$~\cite{mahan:81}. The solution for this Green's function can
be written in the standard form~\cite{berciu:06}: 
\begin{equation} 
G({\mb k}, \omega) =\frac{1}{\omega - \varepsilon_{\bf k} -
\Sigma_{\mr{MA}^{(n)}} (\mb{k}, \omega) + i\eta},
\end{equation}
with the MA$^{(1)}$ self-energy given by:
%\begin{equation}
%\label{eq:Sigma_MA0}
%\Sigma_{\mr{MA}^{(0)}}(\omega) = g A_1(\omega)
%\end{equation}
%and
\begin{equation}
\label{eq:Sigma_MA1}
\Sigma_{\mr{MA}^{(1)}}(\omega) = \frac{g^2 \bar{g}_0(\tilde{\omega})}{1- g
\bar{g}_0(\tilde{\omega})\left[A_2(\omega) - A_1(\omega-\Omega)\right]},
\end{equation}
where 
\begin{equation}
\label{eq:g0}
\bar{g}_0(\omega) = \frac{1}{N}\sum_{\mb{k}} G_0(\mb{k}, \omega)
\end{equation}
is the momentum average over the Brillouin zone of the free propagator
$G_0(\mb{k}, \omega) = (\omega - \varepsilon_{\mb{k}} + i\eta)^{-1}$. 
The functions $A_n(\omega)$ are infinite continued fractions, defined as: 
\begin{eqnarray} \label{eq:An}
A_n(\omega) &=& \frac{n g \bar{g}_0(\omega-n\Omega)}{1- g
\bar{g}_0(\omega-n\Omega) A_{n+1}(\omega)}\\ \nonumber &=&\cfrac{n g
\bar{g}_0(\omega-n\Omega)}{1- \cfrac{(n+1)g^2
\bar{g}_0(\omega-n\Omega)\bar{g}_0(\omega-(n+1)\Omega) }{1-\cdots }}.
\end{eqnarray}
We have also used the short-hand notation $\tilde{\omega} = \omega -
\Omega - gA_1(\omega - 
\Omega)$. As discussed at length elsewhere~\cite{berciu:06}, this MA self-energy is
$\mb{k}$-independent simply because the Holstein model is so
featureless. Indeed,
momentum dependence for $\Sigma(\mb{k}, \omega)$ is obtained for the
Holstein model for MA$^{(n)}$ with $n \ge 2$ (although it is very weak). For
other models with momentum-dependent couplings, a
$\mb{k}$-dependent self-energy is found even at the MA$^{(0)}$
level \cite{goodvin:08}. Further details on the derivation and meaning
of the MA approximation can be found in Ref.~\cite{berciu:06}.  

With an explicit expression for the Green's function of the Holstein
polaron, we calculate the spectral weight 
\begin{equation} \label{eq:spectral}
A(\mb{k}, \omega) = -\frac{1}{\pi} \mr{Im} \, G(\mb{k}, \omega),
\end{equation}
from which we can extract the polaron dispersion, average number of
phonons, effective mass, etc.~\cite{berciu:06}. In our calculations we
employ a small but finite value for $\eta$. This moves the poles of
the Green's function off of the real axis and changes the
$\delta$-peaks of the spectral weight $A(\mb{k},\omega) =
\sum_{\alpha} |\langle \alpha | c_{\mb{k}}^{\dagger} | 0 \rangle |^2
\delta(\omega - E_{\alpha})$ into Lorentzians. In practice, it is
necessary to choose $\eta$ small enough to allow detection of the
Lorentzian peaks in regimes where the \emph{qp} weight is extremely
small. It is also convenient to define the dimensionless coupling
constant $\lambda = g^2/(2d\Omega t)$. We now turn our attention to
the conditions that lead to the formation of an infinitely-lived polaron. 

\section{Results}

Separating the self-energy in terms of its real and imaginary parts,
$\Sigma(\omega) = \Sigma'(\omega) + i \Sigma''(\omega)$, we can write
the spectral function in the following form: 
\begin{equation}
\label{eq:A}
A(\mb{k}, \omega) = \frac{1}{\pi}\frac{\eta -
 \Sigma''(\omega)}{[\omega - \varepsilon_{\mb{k}} -
	\Sigma'(\omega)]^2 + [\eta - \Sigma''(\omega)]^2}. 
\end{equation}
We are interested in the dispersion of the polaron, $E_{\mb{k}}$. This
can be found by tracking the lowest pole in the spectral
weight~\cite{berciu:06}, its energy being the smallest value of $\omega$
which satisfies 
\begin{equation} 
\label{eq:pole}
\omega - \varepsilon_{\mb{k}} = \Sigma'(\omega).
\end{equation}
For this to also be a discrete eigenstate (as opposed to the lower
edge of a continuum), the imaginary part of the self-energy has to
vanish:
\begin{equation} 
\label{eq:im}\left.\Sigma''(\omega)\right|_{\omega=E_{\mb{k}}} = 0.
\end{equation}
The latter requirement is essential, as
eq.~(\ref{eq:A}) does not have singular behavior when the imaginary
part of the self-energy is finite. If eq. (\ref{eq:im}) is satisfied, we
call the solution a true polaron {\em qp}, meaning that it has an infinite
lifetime. If eq. (\ref{eq:im}) is not satisfied, then the
lowest-energy feature is not a discrete state but the
polaron+one-phonon continuum. In this case, we find the energy where
this continuum has a maximum in the density of states (DOS), and call it
a polaron \emph{resonance}. This resonance has 
a finite lifetime proportional to $1/\Sigma''(\omega)$, where $\omega$
is its energy. 

As already mentioned, the onset of the continuum is at $E_0 + \Omega$,
where $E_0$ is the ground state energy of the polaron for
$\mb{k}=0$~\cite{berciu:06}, therefore for $\omega < E_0+\Omega$ it
is guaranteed that $\Sigma''(\omega)=0$. To summarize, we thus look
for the region(s) of parameter space where $E_{\mb{k}}$, the
lowest-energy solution of eq.~(\ref{eq:pole}), satisfies $E_{\mb{k}} <
E_0 + \Omega$. If this condition is satisfied, we have a true polaron
bound state at that momentum. Otherwise, we have a polaron-like
resonance with a finite lifetime lying inside the continuum, as
described above.

\begin{figure}[t]
\onefigure[width=1.0\columnwidth]{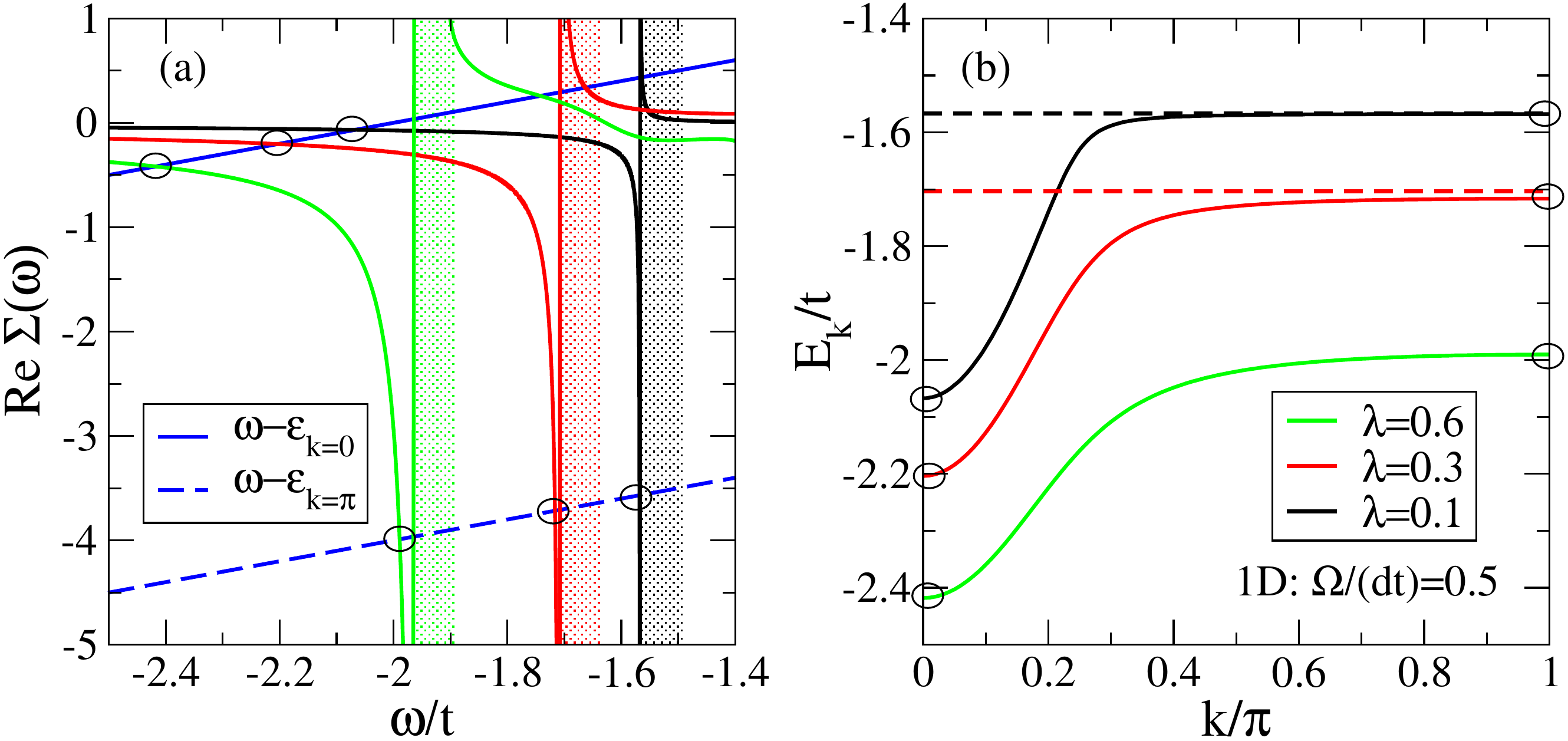}
\caption{(Color online) (a) The real part of the 1D self-energy
 $\Sigma(\omega)$ versus $\omega/t$, and (b) the polaron dispersion
 $E_k$, for effective el-ph coupling strengths $\lambda=0.1, 0.3$
 and 0.6. In (a) we also plot
 $\omega - \varepsilon_k$, for $k=0$ (solid blue line) and $k=\pi$
 (dashed blue line). The intersection of these curves with $\mr{Re}\,
 \Sigma(\omega)$ give the polaron energy eigenvalues, shown by the
 respective circles in (a) and (b). The shaded areas indicate the
 onset of the continuum, \emph{i.e.}, where $\mr{Im} \, \Sigma
 (\omega) \ne 0$. 
\label{fig:1D_Sigma_Ek}}
 \vspace{-0.1cm}
\end{figure}

Obviously, it is the self-energy that controls which of the two
possible cases is encountered. To understand how this occurs, we first
quickly analyze the well-known 1D case, where -- as already mentioned
-- there is always a true polaron {\em qp} state below the
continuum. The reason for this is made obvious by
fig.~\ref{fig:1D_Sigma_Ek}(a), where we show a geometric solution of
eq.~(\ref{eq:pole}) for three different values of the effective
coupling $\lambda$. The real part of the self-energy,
$\Sigma'(\omega)$, goes asymptotically like $\frac{g^2}{\omega}$ as $\omega
\rightarrow -\infty$, so it starts as a negative function which, in
1D, diverges and changes sign at a value marking the onset of the
polaron+one-phonon continuum (above this energy, in the shaded
regions, $\Sigma''(\omega)$ becomes finite). The left-hand side of
eq.~(\ref{eq:pole}) is a straight line -- the limiting $k=0$ and
$k=\pi$ are drawn explicitly -- and a true polaron solution exists if
the two intersect below the onset of the continuum. The divergence of
$\Sigma'(\omega)$ ensures that in 1D there is always such a
solution. Its energy $E_{\mb{k}}$ is shown in
fig.~\ref{fig:1D_Sigma_Ek}(b), where we also plotted the onset of the
continuum, $E_0+\Omega$, for the three cases.

As expected, at very weak couplings and sufficiently large $\mb{k}$, $E_{\mb{k}}$
flattens out just 
below the continuum. Geometrically, we see that this is due to the
fact that for a large region of the Brillouin zone, the intersection
between the two curves occurs very close to the divergence in
$\Sigma'(\omega)$, so the value of the solution $\omega$ varies little and is
asymptotically close to $E_0+\Omega$. The value of $k$ where this
``flattening'' begins is also easy to estimate for $\lambda
\rightarrow 0$: in this limit, to zero order in perturbation theory we
must have $E_{\mb{k}}\approx \varepsilon_{\mb{k}}$, therefore the
``flattening'' must happen for momenta $\mb{k}$ such that $
\varepsilon_{\mb{k}} > \varepsilon_{\mb{0}}+\Omega$. Here, because of
the divergence in the free-electron DOS at the band-edge, arbitrarily
weak coupling to the phonons suffices to repel a bound state below the
continuum. As the coupling $\lambda$
increases, the self-energy curve shifts to lower energies, but the
divergence is always there.

Clearly, the next step is to understand why the self-energy has this
shape and in particular the reason for the divergence in its real
part. If such a divergence is 
always guaranteed to mark the onset of the continuum, then a discrete
polaron state will 
always be found. 

As already discussed, we are actually primarily
interested in the limit of weak el-ph coupling, since at large el-ph
coupling the polaron band is very flat and guaranteed to obey
$E_{\mb{k}}< E_{0} + \Omega$ everywhere in the Brillouin zone. For weak
coupling $g \rightarrow 0$, eq.~(\ref{eq:Sigma_MA1}) gives a
self-energy 
$$
\Sigma(\omega) \approx g^2
\bar{g}_0(\omega-\Omega)+\cdots= \frac{g^2}{N} \sum_{\mb{k}}^{}
G_0(\mb{k}, \omega-\Omega) + \cdots$$
 after using eq.~(\ref{eq:g0}). We
recognize this as being the lowest order diagram contributing to the
self-energy (the Born approximation), in agreement with perturbation
theory. The correction 
$\omega -\Omega \rightarrow \tilde{\omega}$ in the argument of
$\bar{g}_0(\omega)$ will further shift the self-energy towards lower
energies, and is due to corrections from higher order
diagrams. However, this shift cannot be responsible for the appearance
of a singularity in the real part of the self-energy.

\begin{figure}[t]
\onefigure[width=1.0\columnwidth]{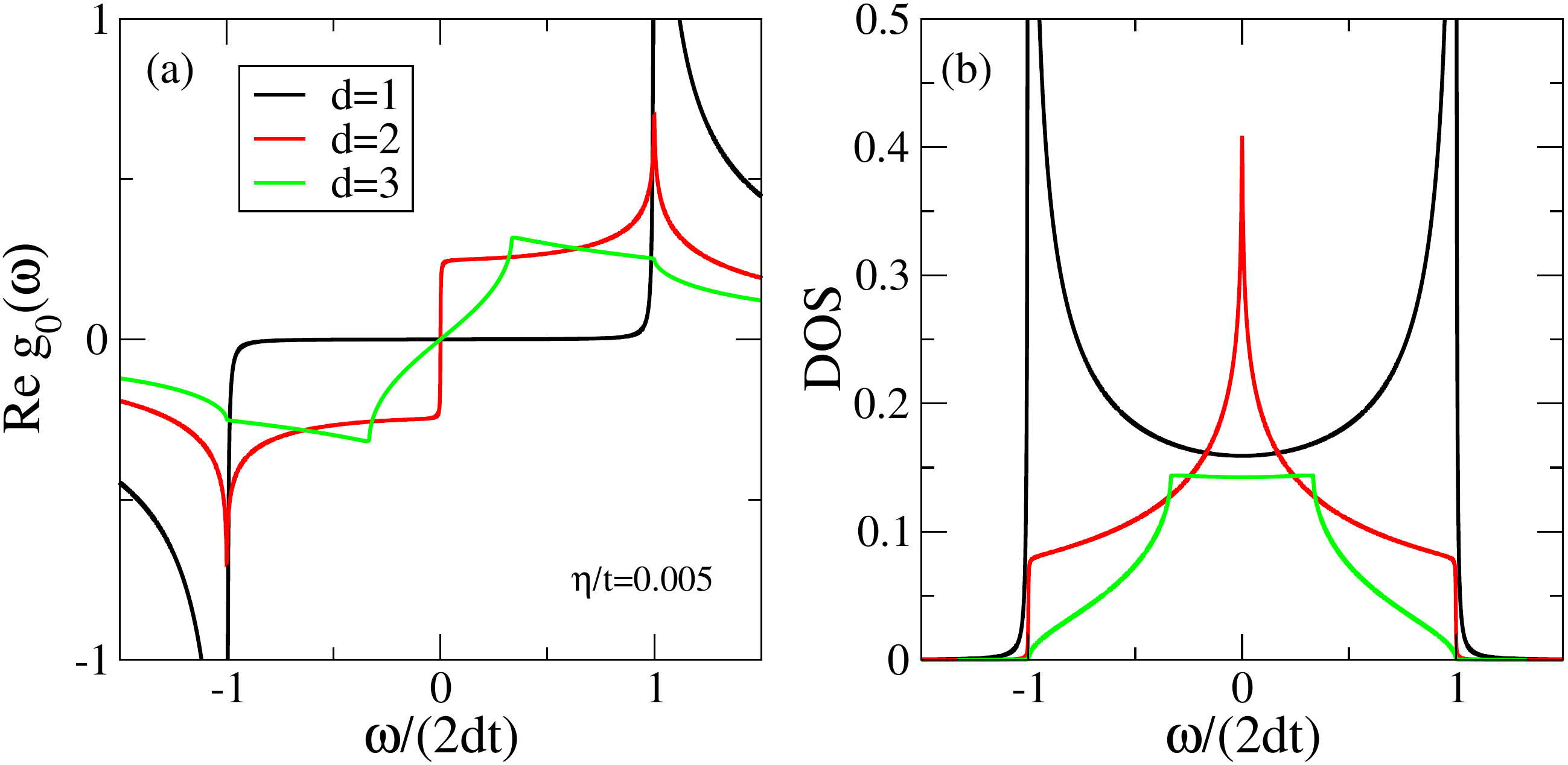}
\caption{(Color online) (a) The real part of $\bar{g}_0(\omega)$ and
 (b) the bare electron DOS $(-1/\pi) \mr{Im}\,
 \bar{g}_0(\omega)$, shown for $d=1,2,3$. A broadening factor of
 $\eta/t = 0.005$ has been used. 
\label{fig:g0_DOS}}
 \vspace{-0.1cm}
\end{figure}

It follows that for the Holstein model (we discuss other models
below) the real part of
$\bar{g}_0(\omega)$ controls the energy dependence of $\Sigma'(\omega)$
for small $\lambda$, and therefore
whether it has a singularity or not. Both the real and imaginary parts
of $\bar{g}_0(\omega)$ are shown for various dimensions in
fig.~\ref{fig:g0_DOS}. Actually, the right panel shows $(-1/\pi)\mr{Im} \,
\bar{g}_0(\omega)$, which from eq.~(\ref{eq:g0}) is seen to be the
bare electron total DOS.

Because $\bar{g}_0(\omega)$ is the momentum average of a retarded
propagator, its real
and imaginary parts are related through
Kramers-Kronig relations. Given that the imaginary part (the DOS) is
finite only inside the free-electron bandwidth $|\omega | \le 2dt$, it
follows that it is the DOS at the band-edge that controls whether the
real part has a singularity at the band-edge energy: if the band-edge
DOS diverges, or at least has a discontinuous jump, then the real part
has a singularity at the band-edge; otherwise it is finite everywhere.

This explains why in 1D we always find a singularity in the real part
of $\bar{g}_0(\omega)$, and therefore in $\Sigma'(\omega)$: it is
well-known that the free electron DOS for nearest-neighbor hopping in 1D has 
van Hove singularities at the band edges. In 2D, the DOS has
 discontinuities at
the band edges. As a result, based on the discussion above, we expect
that in 2D a discrete polaron state must also exist in the entire Brillouin
zone, just like in 1D. However, this is a very weak logarithmic
singularity (as it is due to a discontinuity, not a singularity in the
DOS) and one may expect that it can be easily removed. This is indeed
the case, as we show below where we analyze quasi-2D systems.

\begin{figure}[t]
\onefigure[width=1.0\columnwidth]{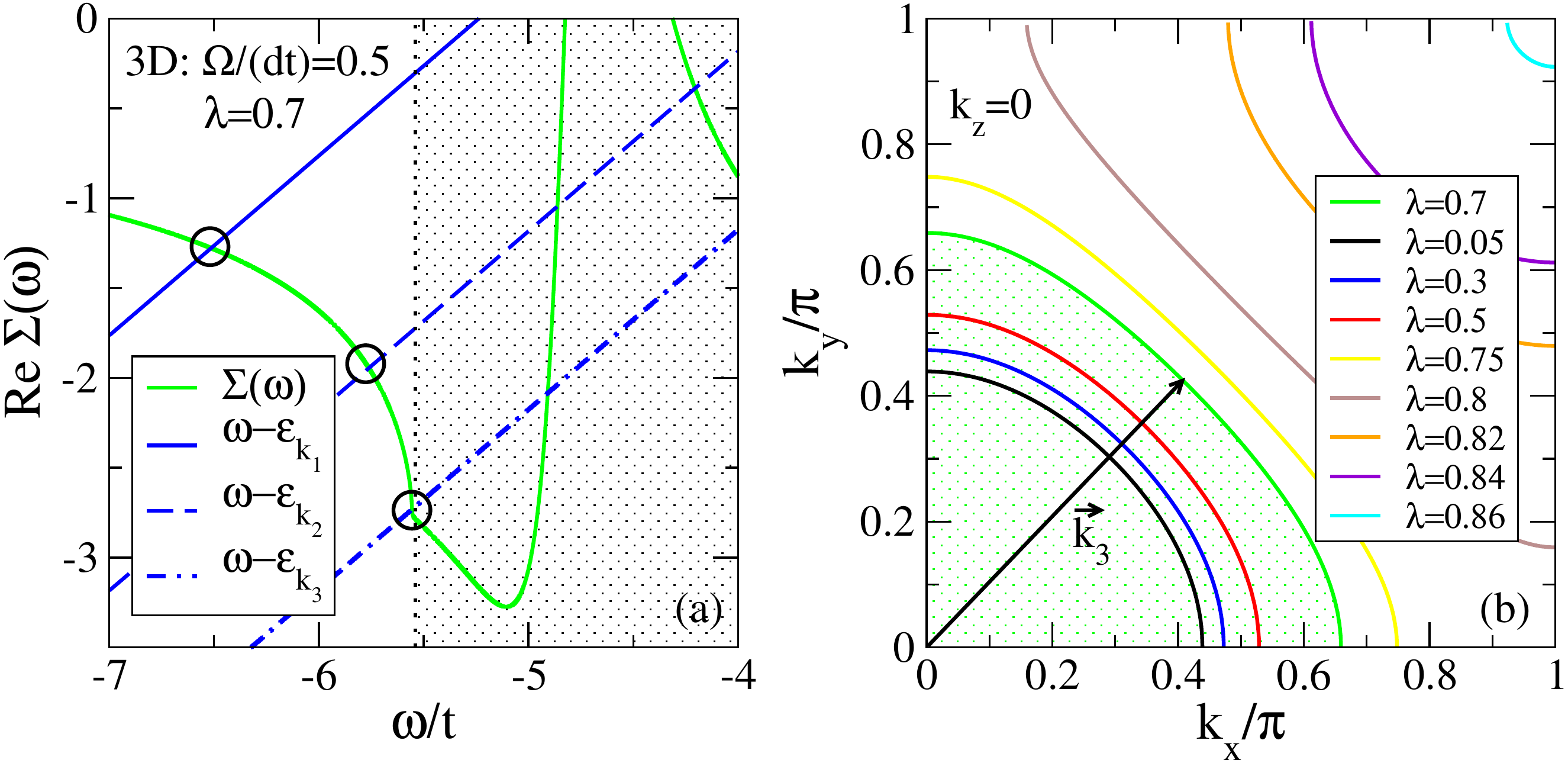}
\caption{(Color online) (a) The real part of the 3D self-energy for an
 intermediate el-ph coupling $\lambda=0.7$. The shaded region shows
 where $\mr{Im}\, \Sigma(\omega) \ne 0$. The straight lines show $\omega -
 \varepsilon_{\mb{k}}$ curves for $\mb{k}_1=\pi(0.2,0.2,0)$,
 $\mb{k}_2=\pi(0.3,0.3,0)$, $\mb{k}_3=\pi(0.43,0.43,0)$. An
 intersection of $\omega - \varepsilon_{\mb{k}}$ and $\mr{Re}\,
 \Sigma(\omega)$ corresponds to a solution of eq.~(\ref{eq:pole}); 
 (b) Contours of the end points for various el-ph coupling
 strengths. The $\lambda=0.7$  contour  is shaded, and the arrow marks
 the end point $\mb{k}_c=\mb{k}_3$ found in (a). 
\label{fig:sigma_kcrit}} 
 \vspace{-0.1cm}
\end{figure}

However, first we focus on the 3D case, where the DOS at the band-edge
is continuous, leading to a real part of $\bar{g}_0(\omega)$, and
therefore low-coupling $\Sigma(\omega)= g^2
\bar{g}_0(\tilde{\omega})$, that is finite everywhere. 
The geometric solution of eq.~(\ref{eq:pole}) in this case is
illustrated in the left panel of fig.~\ref{fig:sigma_kcrit}, for an
intermediate effective coupling $\lambda=0.7$. The low-energy part of
$\Sigma'(\omega)$ still resembles the real part of
$\bar{g}_0(\omega)$, although due to higher order diagrams, the
shape is somewhat distorted. The differences are more considerable at
higher energies (this issue is revisited below). The onset of
the continuum at $E_0+\Omega$ is marked by a
discontinuity in the slope of 
$\Sigma'(\omega)$ (shaded area). Because $\Sigma'(\omega)$ does not diverge below
this energy, it is now apparent that eq.~(\ref{eq:pole}) only has
 solutions with $E_{\mb{k}}< E_0+\Omega$ up to a critical
value $\mb{k}_c$ which marks the ``end point'' in the polaron dispersion
\cite{Prok}. For $\lambda=0.7$ and along the $(1,1,0)$ direction,
the critical value is at $\mb{k}_c=\pi(0.43,0.43,0)$, as shown
in fig.~\ref{fig:sigma_kcrit}(a). 

Repeating this along various directions allows us to identify the
surface described by the ``end points'' $\mb{k}_c$ in the Brillouin
zone, which marks the separation between the region near the
$\Gamma$ point where an infinitely-lived quasiparticle exists, and the
region near the edges of the Brillouin zone where only a resonance
with a finite lifetime appears. The end points contours in a
quadrant of the $k_z=0$ plane are shown in fig.~\ref{fig:sigma_kcrit}(b), for
various values of $\lambda$. The arrow marks
the particular $\mb{k}_c$ found in fig.~\ref{fig:sigma_kcrit}(a). 

For a given value of the phonon mode,
the smallest surface of end points is
for $\lambda \rightarrow 0$. By analogy with a previous discussion,
in this case we can estimate it to be given by
$\varepsilon_{\mb{k}_c} \approx 
\varepsilon_{0}+\Omega$, since for $\lambda \rightarrow 0$,
$E_{\mb{k}} \approx \varepsilon_{\mb{k}}$. It follows that for any
$\Omega >0$, there is a vicinity of the $\Gamma$ point where a true
polaron exists. In general, however, there are regions in the
Brillouin zone that do not satisfy this condition for small
$\lambda$. Indeed, since the 
maximum value of $\varepsilon_{\mb{k}}$ is $6t$ while
$\varepsilon_{0}=-6t$, we estimate that for 
any $\Omega < 12 t$ there exist such regions of momenta without an
infinitely lived polaron state. $\Omega < 12 t$ is likely to
hold in most materials, therefore such a case is rather
typical. 
If $\Omega$ is not too large, then in the limit $\lambda \rightarrow
0$ the surface of end points is spherical, since
at low energies the free-electron dispersion can be approximated by a
parabola. As either $\Omega$ or $\lambda$ increase, the shape will
expand and distort, as seen in fig.~\ref{fig:sigma_kcrit}(b) for
increasing $\lambda$. For any value of $\Omega<12t$, there is a
critical value $\lambda_c>0$ above which an infinitely-lived polaron
solution appears in the entire Brillouin zone. For $\Omega=0.5t$,
fig.~\ref{fig:sigma_kcrit}(b) reveals that this $\lambda_c$ is
just above 0.86.

\begin{figure}[t]
\onefigure[width=1.0\columnwidth]{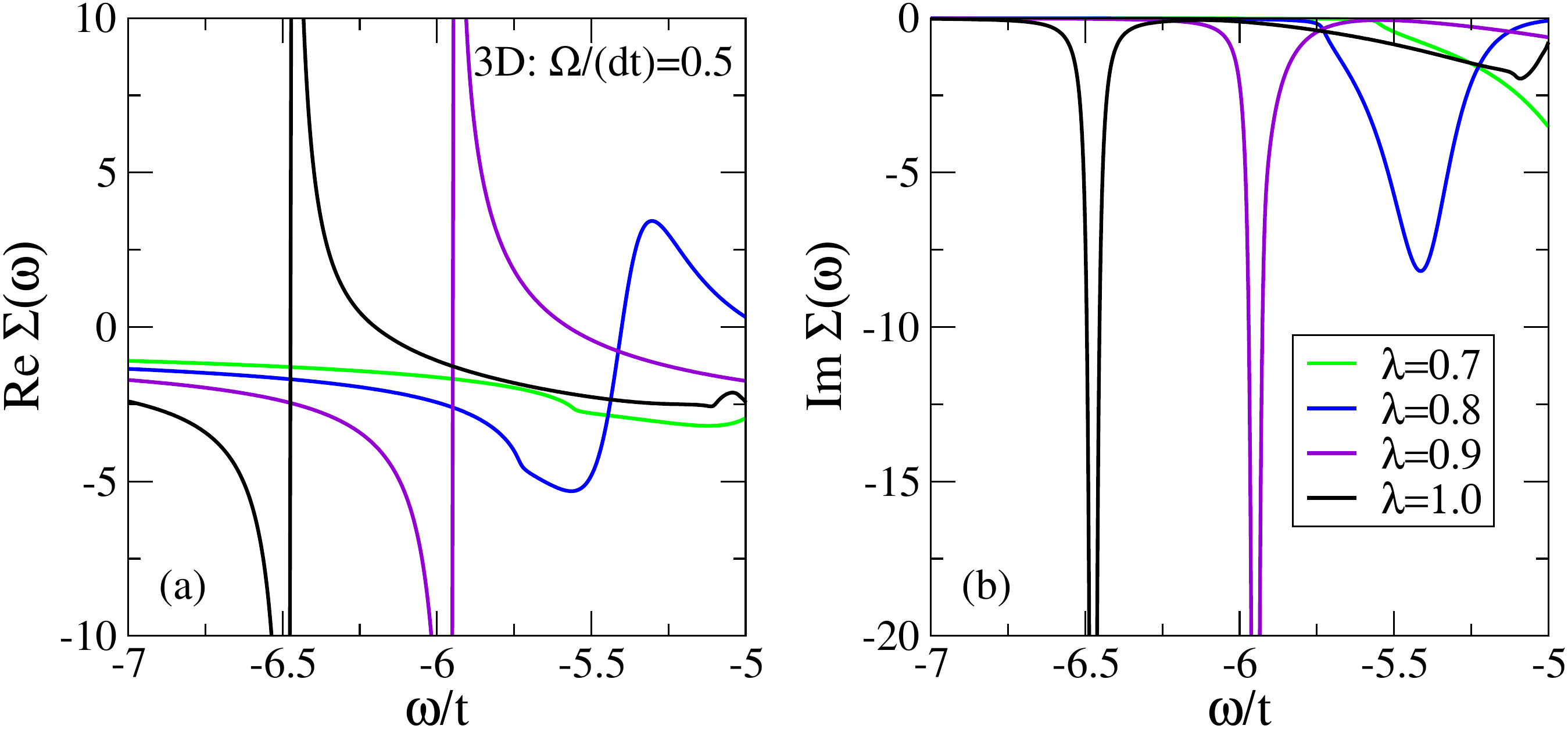}
\caption{(Color online) The (a) real and (b) imaginary parts of the 3D
 self-energy, for increasing el-ph coupling strengths. The
 self-energy becomes divergent for large enough coupling. 
\label{fig:sigma1_lambda}}
 \vspace{-0.1cm}
\end{figure}

As already discussed, the fact that a true polaron solution exists
everywhere in the Brillouin zone at strong(er) coupling is not
surprising, because the significant increase in the polaron effective
mass guarantees that eventually the whole polaron band will fit below the
continuum. In terms of the geometrical solution, what happens is
revealed in fig. \ref{fig:sigma1_lambda} which shows that for large
enough $\lambda$, the real part of the self-energy does eventually
gain a singularity at low-energies. This divergence is a higher order
effect, not coming from a divergence/discontinuity in the DOS, but due
to the structure of the self-energy itself.

Eq.~(\ref{eq:Sigma_MA1}) reveals that at large enough
el-ph coupling $g$, the denominator itself can vanish giving rise to
a different type of 
divergence in the self-energy. The appearance of this divergence
explains the phenomenology related to $\lambda_c$ (for $\Omega=0.5t$,
this divergence is first observed at $\lambda_c=0.87$). Physically, it
is linked to the formation of the second bound state
\cite{bonca:99,berciu:06}, which is known 
to form below the polaron+one-phonon continuum as the crossover
towards the small polaron regime is approached for $\lambda \sim 1$.

For completeness, we mention that as the
coupling increases further, more and more divergences appear in
$\Sigma'(\omega)$ due to the structure of the continuous fractions,
see eq.~(\ref{eq:An}). These additional
divergences occur at higher and higher 
energies that are well above the onset of the continuum, thus they are not
relevant for the problem we
consider here (they are linked to the crossover towards a Lang-Firsov
type of spectrum expected in the limit $\lambda \rightarrow \infty$). 

\begin{figure}[t]
\onefigure[width=1.0\columnwidth]{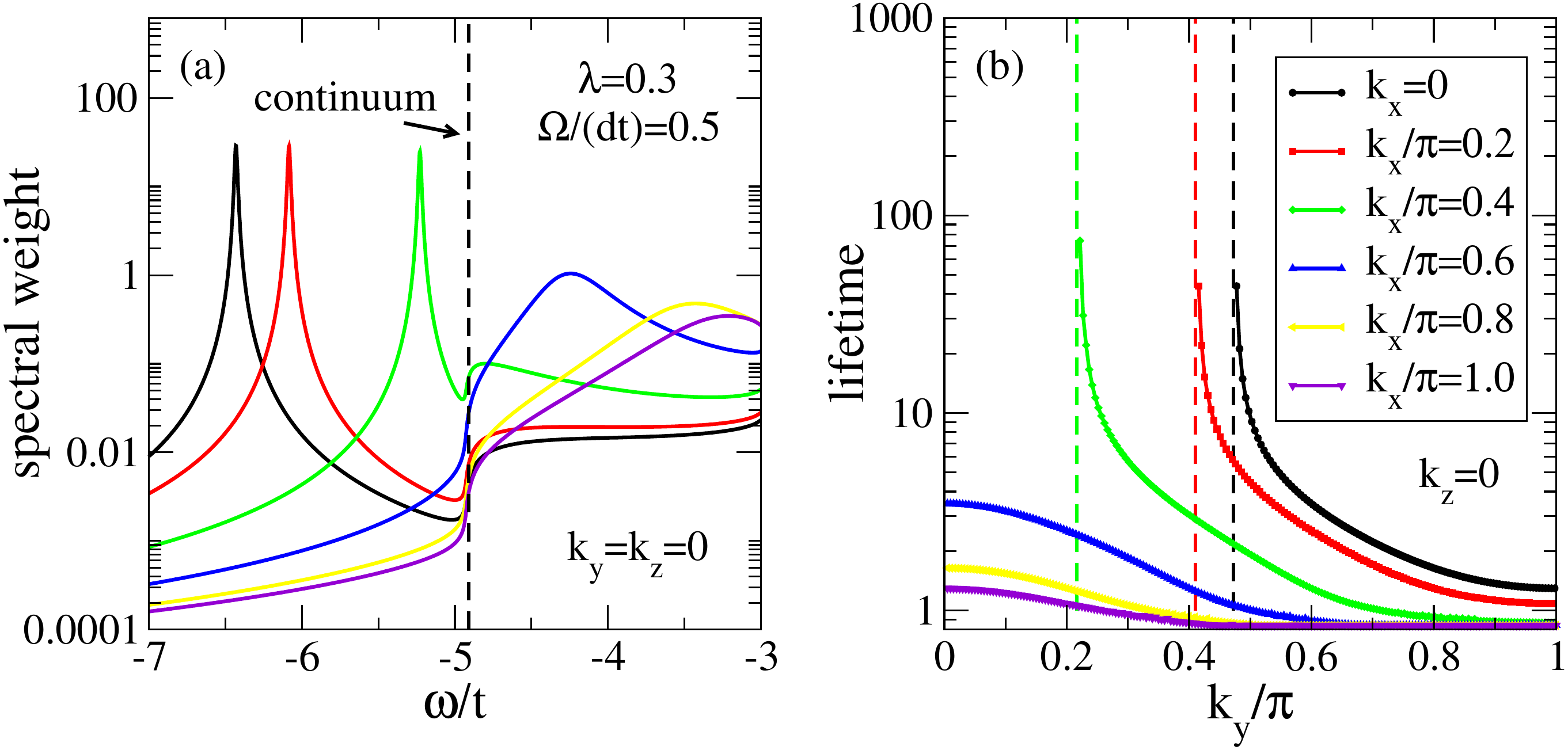}
\caption{(Color online) (a) The spectral weight $A(\mb{k},\omega)$
 versus $\omega$ with $k_y=k_z=0$, and
 $k_x/\pi=0,0.2,0.4,0.6,0.8,1.0$. The dashed line denotes the onset
 of the continuum. With the parameters shown, there is no longer a
 polaron {\em qp} for $k_x/\pi>0.46$, but a polaron-like resonance. (b)
 The lifetime of the polaron-like resonance as a function of
 momentum. The dashed lines denote the transition to an infinitely
 lived polaron state. 
\label{fig:lifetime}}
 \vspace{-0.1cm}
\end{figure}

What happens for $\lambda < \lambda_c$ as we sweep through $\mb{k}_c$ is
shown in fig.~\ref{fig:lifetime}(a), where we plot the spectral weight
$A(\mb{k},\omega)$ vs. $\omega$, for several values of $\mb{k}$. The
onset of the continuum is marked by the dashed vertical line. For
momenta below $\mb{k}_c$, we see a Lorentzian peak below the
continuum. Its width is controlled by $\eta$. This is the discrete
level whose energy is $E_{\mb{k}}$. As $\mb{k}$ increases,
$E_{\mb{k}}$ approaches the continuum. Unlike in 1D and 2D cases,
however, where the singularity/discontinuity in the DOS ensured that a
discrete level is always pushed below the continuum, here we see
that above $\mb{k}_c$, the low-energy feature is the continuum,
with a broad maximum that disperses towards higher energies while
broadening even more as $\mb{k}$
increases -- this is the polaron {\em resonance}. Its finite
lifetime, equal to $1/\Sigma''(\omega)$ where $\omega$ is its
energy, is displayed in fig.~\ref{fig:lifetime}(b) along various cuts
in the Brillouin zone. As expected, it diverges as $\mb{k}_c$ is
approached from above, and it decreases very fast towards the edges of
the Brillouin zone. This is a $T=0$ calculation, so this finite lifetime is an
intrinsic effect. As a technical note, in calculating it we
decrease $\eta$ until changing it any further has no effect on the
lifetime.

\begin{figure}[t]
\onefigure[width=1.0\columnwidth]{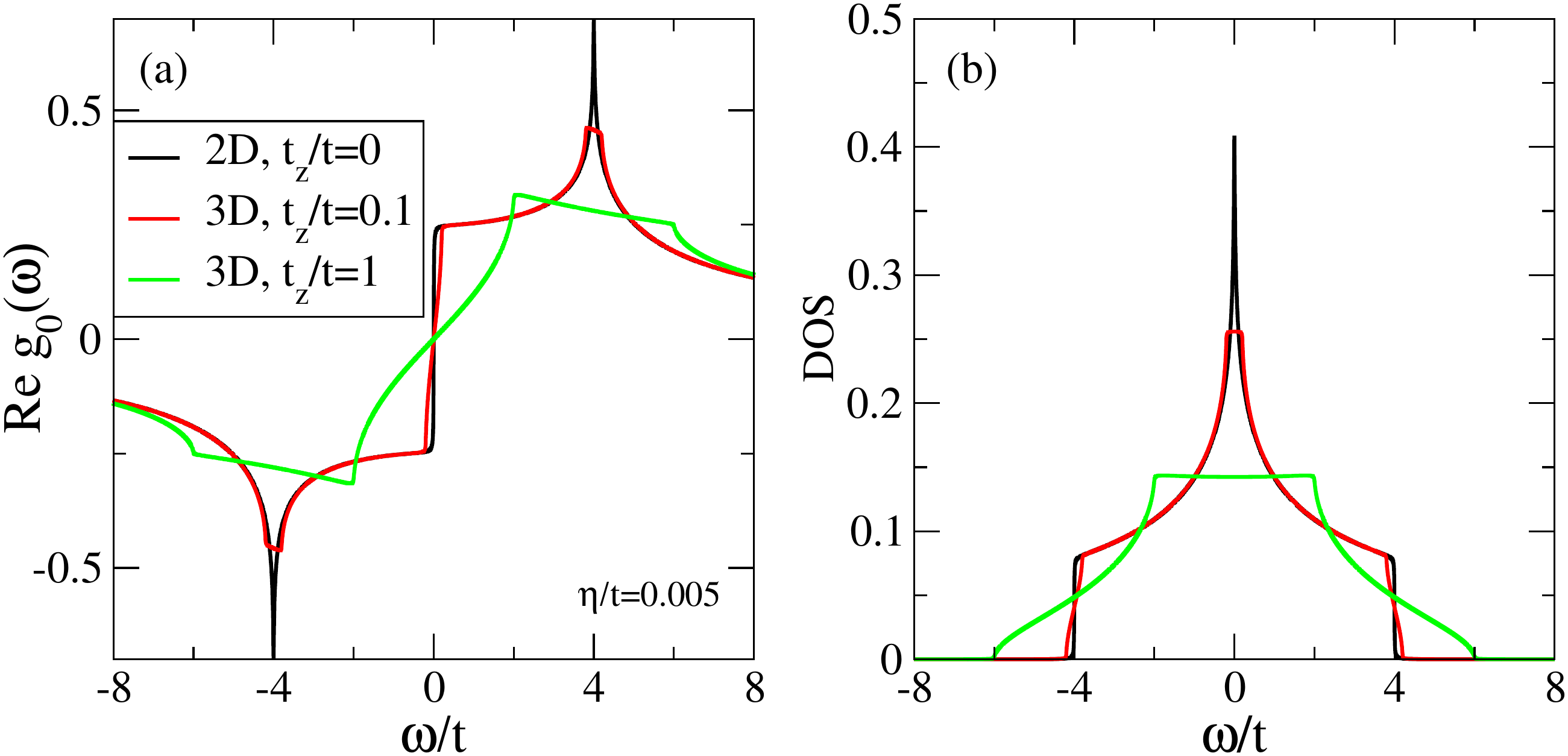}
\caption{(Color online) Anisotropic hopping - transition from 2D to 3D DOS.
\label{fig:g0_aniso}}
 \vspace{-0.1cm}
\end{figure}

Based on all these results, we can now quickly analyze the 2D
case. For a purely planar 2D hopping model, the well-known
discontinuity in the DOS at the band-edge guarantees that a low-energy
singularity always exists in $\Sigma'(\omega)$, see
fig.~\ref{fig:g0_DOS}. As a consequence, a true polaron is guaranteed
to form for all $\mb{k}$ at any $\lambda$, just like in 1D. However,
most layered materials are in fact quasi-2D, because there is some
small anisotropic hopping between different planes. The effect of a
small $t_z=0.1 t$ on the $\bar{g}_0(\omega)$ function is shown in
fig.~\ref{fig:g0_aniso}. Any $t_z\ne 0$ will smooth out the band-edge
discontinuity in the DOS, and thus remove the weak singularity in the
real part of $\bar{g}_0(\omega)$ and therefore in $\Sigma'(\omega)$ at
weak coupling. As a result, quasi-2D systems behave similarly to 3D
systems: for weak-to-intermediate coupling, we expect that a surface
of end points separate the region near $\Gamma$ where a true
polaron quasiparticle forms, from a region near the Brillouin zone
edges where only a finite lifetime resonance appears. The detailed
shape of this surface and the critical couplings $\lambda_c$ above
which a true polaron exists everywhere will, of course, depend on the
particular parameters of the problem. However, the reasons we
uncovered here to explain this phenomenology are robust and therefore
such surfaces of end points should be quite generic in quasi-2D
and 3D materials with not too large el-ph coupling.

The existence of such surfaces of end points, especially in
layered systems, should be directly visible using Angle Resolved
Photo-Emission Spectroscopy \cite{damascelli:03}. This technique
measures the spectral
weight and should be able to identify significant changes in its
shape, like those shown in fig.~\ref{fig:lifetime}(a). Very different
temperature dependence of features in different regions of the
Brillouin zone would also be consistent with such phenomenology,
because the true {\em qp} peaks should only exhibit thermal
broadening, while the finite-lifetime resonances have an intrinsic
lifetime. Of course, many other properties should be strongly
influenced by the existence of such a surface of end points, especially if the
Fermi surface is not too far from it (this statement assumes
that interactions between polarons, either due to direct interactions
between the particles or mediated through exchange of bosons between
the two polaronic clouds, do not qualitatively change this
phenomenology. Whether this is the case or not is a very interesting
question which still awaits resolution).

To summarize, for a Holstein model and weak-to-intermediate coupling
one can deduce easily whether an infinite lifetime
polaron {\em qp} is expected everywhere in the Brillouin zone or not
simply from knowing whether the DOS of the free
electrons is continuous or not at the band-edge. To pinpoint the
surface of end points, more detailed calculations like the ones shown
here are needed. For more complicate
polaron models, with a particle-boson coupling that depends on either
(or both) the particle and boson momentum, and also with a possibly
dispersing boson mode, one can certainly gain
intuition about what is likely to happen from studying the Born approximation
$$
\Sigma(\mb{k},\omega) \approx \frac{1}{N} \sum_{\mb{q}}^{} |g_{\mb{k},\mb{q}}|^2
G_0(\mb{k}-\mb{q}, \omega - \Omega_{\mb{q}})
$$
and whether it generically predicts infinitely-lived quasiparticles
for all $\mb{k}$ or not (if there are no end points at weak coupling,
they are very unlikely to appear at stronger couplings for reasons already
discussed). If either $g_{\mb{k},\mb{q}}$ or
$\Omega_{\mb{q}}$ have strong momentum dependence, then
$\Sigma(\mb{k},\omega)$ may acquire a sufficiently strong
$\mb{k}$-dependence to modify the conditions found above for the
Holstein model. In particular, it is not apriorily obvious that a
singularity in $\Sigma'(\mb{k},\omega)$ is absolutely needed to
guarantee a discrete low-energy solution everywhere in the Brillouin
zone, if its momentum is strong enough to shift the curve around significantly.

%%%%%%%%%%%%%%%%%%%%%%%%%%%%%%%%%%%%%%%%%%
\acknowledgements
This work was supported by NSERC and CIfAR. We thank N. Prokof'ev for
suggesting the problem, and G. A. 
A. Sawatzky for many useful discussions. 

%%%%%%%%%%%%%%%%%%%%%%%%%%%%%%%%%%%%%%%%%%


\begin{thebibliography}{99}

%\bibitem{landau:33}
% \Name{Landau L. D.}
% \REVIEW{Phys. Z. Sowjet}{3}{1933}{644}.

%Manganites, cuprates, fullerenes (organic materials)
\bibitem{salamon:01}
\Name{Salamon M.B. \and Jaime M.}
\REVIEW{Rev. Mod. Phys.}{73}{2001}{583}.

\bibitem{shen:04}
\Name{Shen K.M. \etal}
\REVIEW{Phys. Rev. Lett.}{93}{2004}{267002}.

\bibitem{hengsberger:99}
\Name{Hengsberger M., \etal}
\REVIEW{Phys. Rev. Lett.}{83}{1999}{592}.

\bibitem{matsui:10}
\Name{Matsui H. \etal}
\REVIEW{Phys. Rev. Lett.}{104}{2010}{056602}.

\bibitem{holstein:59}
\Name{Holstein T.}
\REVIEW{Ann. Phys. (N.Y.)}{8}{1959}{325}.
 
\bibitem{fehske:07}
 \Editor{Alexandrov A.S.}
 \Book{Polarons in Advanced Materials}
 \Publ{Canopus, Bath/Spring-Verlag, Bath}
 \Year{2007}.

\bibitem{Prok}
\Name{Prokof'ev N.V. \and Svistunov B.V.}
\REVIEW{Phys. Rev. Lett.}{81}{1998}{2514}.
 
  \bibitem{romero:99}
 \Name{Romero A.H., Brown D.W., \and Lindenberg K.}
 \REVIEW{Phys. Rev. B}{60}{1999}{14080}.
 
 \bibitem{ku:02}
 \Name{Ku L.-C., Trugman S., \and Bonca J.}
 \REVIEW{Phys. Rev. B}{65}{2002}{174306}.

\bibitem{berciu:06}
 \Name{Berciu M.}
 \REVIEW{Phys. Rev. Lett.}{97}{2006}{036402};
 \Name{Goodvin G.L., Berciu M., \and Sawatzky G.A.}
 \REVIEW{Phys. Rev. B}{74}{2006}{245104}; 
 \Name{Berciu M. \and Goodvin G.L.}
 \REVIEW{Phys. Rev. B}{76}{2007}{165109}.
 
\bibitem{kornilovich:99}
\Name{Kornilovich P.E.}
\REVIEW{Phys. Rev. B}{60}{1999}{3237}.
 
 \bibitem{goodvin:08}
 \Name{Goodvin G.L. \and Berciu M.}
 \REVIEW{Phys. Rev. B}{78}{2008}{235120};
 \Name{Berciu M. \and Fehske H.} (unpublished).
 %\REVIEW{Phys. Rev. B}{xx}{2010}{xxxxxx}.

\bibitem{berciu:10}
 \Name{Berciu M., Mishchenko A.S. \and Nagaosa N.}
 \REVIEW{Euro. Phys. Lett.}{89}{2010}{37007};
 \Name{Goodvin G.L., Covaci L. \and Berciu M.}
 \REVIEW{Phys. Rev. Lett.}{103}{2009}{176402}.

 \bibitem{goodvin:10}
 \Name{Goodvin G.L. \and Berciu M.} (unpublished).
 %\REVIEW{unpublished}a{xx}{xxxx}{xxxxxx}.
 
\bibitem{mahan:81}
 \Name{Mahan G.D.}
 \Book{Many particle physics}
 \Publ{Plenum, New York}
 \Year{1981}.
 
 \bibitem{bonca:99}
 \Name{Bonca J., Trugman S. A.  \and Batistic I.}
 \REVIEW{Phys. Rev. B}{60}{1999}{1633}.
 
\bibitem{damascelli:03}
 \Name{Damascelli A., Hussain Z., \and Z.-X. Shen}
 \REVIEW{Rev. Mod. Phys.}{75}{2003}{473}.
 
 

 
\end{thebibliography}
\end{document}